# Helical Edge States and Quantum Phase Transitions in Tetralayer Graphene

Shi Che[1*], Yanmeng Shi[2*], Jiawei Yang[1], Haidong Tian[1], Ruoyu Chen[1], Takashi Taniguchi[3], Kenji Watanabe[3], Dmitry Smirnov[4], Chun Ning Lau[1†], Efrat Shimshoni[5], Ganpathy Murthy[6], Herbert A. Fertig[7‡]

[1] Department of Physics, The Ohio State University, Columbus, OH 43210
[2] Department of Physics and Astronomy, University of California, Riverside, Riverside, CA 92521
[3] National Institute for Materials Science, 1-1 Namiki, Tsukuba, Ibaraki 305-0044, Japan
[4] National High Magnetic Field Laboratory, Tallahassee, FL 32310
[5] Department of Physics, Bar-Ilan University, Ramat Gan 52900, Israel.
[6] Department of Physics, University of Kentucky, Lexington, KY 40506
[7] Department of Physics, Indiana University Bloomington, Bloomington, IN 47405

**Abstract**
Helical conductors with spin-momentum locking are promising platforms for Majorana fermions. Here we report observation of two topologically distinct phases supporting helical edge states in charge neutral Bernal-stacked tetralayer graphene in Hall bar and Corbino geometries. As the magnetic field $B_\perp$ and out-of-plane displacement field $D$ are varied, we observe a phase diagram consisting of an insulating phase and two metallic phases, with 0, 1 and 2 helical edge states, respectively. These phases are accounted for by a theoretical model that relates their conductance to spin-polarization plateaus. Transitions between them arise from a competition among inter-layer hopping, electrostatic and exchange interaction energies. Our work highlights the complex competing symmetries and the rich quantum phases in few-layer graphene.

Helical conductors, systems that have no bulk conduction but support dissipationless conducting states at their edges, may be engineered to realize Majorana statistics for quantum computation[1-4]. Underlying these remarkable systems are the non-trivial topology of electronic structure in the bulk, arising in part from the states associated with a valence band that are energetically raised above those of a conduction band. At the system boundary, this inversion is relaxed, leading to crossing of "hole-like" states of the valence band with "electron-like" states of the conduction band. In helical conductors, these states carry different spin quantum numbers, protecting the crossing and preventing a gap from opening in the spectrum of edge states. Such a band inversion is typically induced by large spin-orbit coupling in topological insulator materials at zero magnetic field[5-7]. Alternatively, they may manifest in semimetals with coexisting electron and hole pockets in the quantum Hall regime, though the mobility of such systems realized in traditional semiconductors is relatively low[8]. These systems are typically not tunable in situ, and helical conduction is only achieved over a narrow range of parameters.

The advent of few-layer graphene provides an alternative platform to realize helical edge states: as semi-metals, they host electron and hole states coexisting near the charge neutrality point (CNP)[9-15], while affording high mobility transport. Helical edge states are observed in monolayer graphene at the charge neutrality point (CNP) in the presence of large in-plane magnetic fields[16], in bilayer graphene in displacement fields[17], and in trilayer graphene in modest magnetic fields[12].

---

† Email: lau.232@osu.edu
‡ Email: hfertig@indiana.edu

Here we report observation of quantum phases supporting helical edge states in tetralayer graphene (4LG) at the CNP in a large magnetic field, arising from the competing effects of interlayer coherence, electrostatic polarization and exchange interaction. As the interlayer potential and magnetic field varies, we observe a series of quantum transitions among the phases that host 2, 1 and 0 helical edge states on each edge, with quantum critical phase boundaries that move with parallel magnetic field. Our work highlights the complex competing symmetries in few-layer graphene and the rich quantum phases in this seemingly simple system. As 4LG is a few-layer graphite system that is tunable by gate, our observations may also be relevant to the highly resistive state observed in bulk graphite[18-23], whose underlying nature remains controversial to date.

The low energy electron bands of 4LG consist of two bilayer graphene (BLG)-like bands with light (m) and heavy (M) effective masses (Fig. 1a), which intersect and hybridize near the CNP due to next-nearest interlayer hoppings (Fig. S1a). The unique band structure and high tunability give rise to multiple Lifshitz transitions and multi-band transport when the magnetic field is absent or small[24-27]. We note that inversion (and thus valley) symmetry is always preserved in the absence of external fields.

We perform magneto-transport measurement on dual-gated 4LG devices with Hall bar and Corbino geometris[28] (Fig. 1b-c), with independently controlled charge density $n$ and out-of-plane displacement field $D$. The experiments were performed in a $^3$He cryostat employing lock-in techniques. The devices' field-effect mobility ranges from 30,000 to 100,000 cm$^2$/Vs. All measurements are taken at $T$=0.3 K unless specified otherwise.

Fig. 1d displays the longitudinal resistance $R_{xx}=V_{23}/I_{14}$ of a Hall bar device (H1) vs. $n$ and perpendicular magnetic field $B_\perp$ at $D$=0. Here the subscript numbers denote the terminals of voltage and current probes as indicated in Fig. 1b. A well-defined Landau fan is observed. At $n$=0 and relatively low field, the device is highly resistive with $R_{xx}$ ranging from 100 kΩ to 2 MΩ, similar to those observed in monolayer and bilayer graphene[16, 29-35]. However, when $B_\perp$ exceeds ~30 T, $R_{xx}$ drops precipitously to ~ 7 kΩ (Fig. 1e). This dramatic decrease in resistance has not been observed in other graphene systems, suggesting a field-induced insulator-metal transition.

To explore the electronic phases of the $\nu$=0 QH state, we measure $R_{xx}$ vs. $B_\perp$ and $D$, while maintaining overall charge neutrality. At large magnetic field, a striking phase diagram emerges (Fig. 2a). Guided by the boundaries between dramatically changed $R_{xx}$, we identify three different phases. The brown region, identified as phase I, indicates a highly insulating state ($R_{xx}$ >50 kΩ). It develops at a moderate $B_\perp$ (~6 T) and persists over the entire range of $D$ (up to ±600 mV/nm) until $B_\perp$=22 T. This insulating state transitions abruptly to conductive regions with $R_{xx}$~7-12 kΩ, or equivalently, conductance that is approximately 2-4 times the conductance quantum $G_Q=e^2/h$. Interestingly, the phase boundary that separates the insulating and conductive states is not monotonic in the $B_\perp$-$D$ plane, but has a "Σ" shape: the transition at $B_\perp$=30 T occurs at $D$=0, and $B_\perp$=22 T at $D$=280 mV/nm. Within the conductive regions, phase II (blue) has the lower resistance, with $R_{xx}$ ranging from 4 to 8 kΩ. It starts at the transition point from phase I at $B_\perp$~30 T and $D$=0, its phase boundary expanding linearly with $B_\perp$, and re-appears at larger $D$. Lastly, the green region dominating the conducting regime at moderately low $D$ has resistance ~12 kΩ, and is identified as phase III. Here we identify phases II and III as distinct phases, due to their conductances that are nearly quantized at low temperature to ~ $4e^2/h$ and $2e^2/h$, respectively (Fig. 2d inset). Transitions between various phases as a function of $D$ and $B_\perp$ are illustrated by the line cuts in Fig. 2b-d. The point at $B_\perp =30$ T and $D=0$ constitutes a quantum critical point, apparently adjoining all three phases, which we denote $B_{cp}$. The overall phase diagram is observed in 4 Hall bar devices, and a similar partial phase diagram is mapped in additional 5 Hall bar devices.

The insulating (metallic) nature of phase I (II and III) is further confirm by their temperature dependence. As $T$ increases, $R_{xx}$ of phase I decreases from ~80 kΩ to ~40 kΩ; in contrast, in phase II, $R_{xx}$ rises from ~6 kΩ at base temperature to become nearly as resistive as phase I at 40 K (Fig. 2b). Similarly, at $B_\perp$=25 T, phase II is indicated by the resistance valleys centered at $|D|$~280 mV/nm, where $R_{xx}$ increases from ~5 kΩ to ~40 kΩ as $T$ is raised 20 K (Fig. 2c). At $B_\perp$=34 T, $R_{xx}$ of the resistive state for $|D|$>600 mV/nm drops with increasing temperature, signifying its insulating nature.

A central question in ascertaining the natures of these phases is whether conduction therein occurs via bulk or edge state transport, as $R_{xx}$ in Hall bar devices contains contributions from both mechanisms. To address this question, we fabricated dual-gated devices with Corbino geometry (Fig. 1c), in which no physical edges connect the electrodes and therefore probe *only bulk transport*. The conductance *difference* between Corbino and Hall bar devices then originate solely from edge states. Fig. 3a plots the two-terminal conductance $G_{\text{Corbino}}(B_\perp, D)$ from a Corbino device (C1). Phase I is insulating in both the Hall bar and Corbino devices; the absence of a gap transition point suggests a first-order transition, in agreement with Hartree-Fock calculations (see Fig. 4 and associated discussion). Interestingly, $G_{\text{Corbino}}(B_\perp, D)$~0 while $R_{xx}$ ~$h/4e^2$ for phase II, indicating an insulating bulk and high edge state conductance. In phase III, a somewhat higher $G_{\text{Corbino}}$~$0.5e^2/h$ suggests that bulk excitations have a relatively small gap; nevertheless, much of the conductance in phase III is also contributed by edge states. Similar phase diagrams are observed in 3 Corbino devices.

To better visualize the edge conductance, we note that the longitudinal conductance of the Hall bar device is a sum of the edge and bulk conduction, $G_{Hall}=G_{edge}+\sigma_{bulk}(W/L)$, while $G_{Corbino}$ arises sole from the bulk, $G_{Corbino} = \frac{2\pi}{\ln\left(\frac{b}{a}\right)}\sigma_{bulk}$ (here $\sigma_{bulk}$ is the conductivity of the bulk, $W/L$ is the aspect ratio of the Hall bar channel, $b$ is the distance between the two contacts, and $a$ is the radius of the inner contact). Thus, the edge conduction is given by $G_{edge} = G_{Hall} - \frac{W}{L}\frac{\ln\left(\frac{b}{a}\right)}{2\pi}G_{Corbino}$. From device geometries, we estimate that $\frac{W}{L}\frac{\ln\left(\frac{b}{a}\right)}{2\pi} \sim 0.1$. Combining data from devices H1 and C1, we plot $G_{edge}(B_\perp, D)$ in Fig. 3b, thus explicitly demonstrating that the edge conduction approaches $4e^2/h$ in phase II and $2e^2/h$ in phase III. In addition, non-local measurements also yield conductance values that agree with Landauer-Buttiker formalism for 2 and 1 helical edge states, respectively[36]. Taken together, these data sets unambiguously establish edge state conduction in phase II and III.

Lastly, we investigate how the phases are affected by an in-plane magnetic field $B_\parallel$. Fig. 3c displays $R_{xx}(B_\perp, D)$ of H1 that is tilted at an angle $\theta$=49°. The general features resemble that in Fig. 2a with $B_\parallel$=0; however, the entire phase boundaries shift towards lower $B_\perp$. For instance, the transition between phase I and II at $D$=0 now takes place at $B_\perp$=27 T. Such Zeeman-induced reduction of the critical magnetic field $B_{\perp c}$ indicates spin-ordering. A detailed Zeeman energy dependence of transition at $n=D=0$ from phase I to II for another Hall bar device H2 is plotted in Fig. 3d, showing the quantum critical point $B_{\perp cp}$ moves to lower values as larger $B_\parallel$ is applied. In fact, $B_{\perp cp}$ decreases linearly with total applied field $B_t$, suggesting that the transition point is linearly dependent on the Zeeman energy (Fig. 3d).

To understand the phase diagram and the origin of the metallic states, we first calculate the LL spectrum of 4LG using a non-interacting $k \cdot p$ continuum model[26, 27]. The hopping parameters are extracted by fitting calculated spectra to experimentally observed LL crossing points, and are consistent with previous reports[27, 37]. Fig. 4a displays the calculated LL spectrum at $D=0$, where each LL is valley- and spin-degenerate, labeled by spin (↑ and ↓), valley (K and K'), orbital ($N=-1$ and 0), and bands with heavy (M) and light (m) masses. Red and blue curves denote electron-like and hole-like LLs, respectively.

Of most relevance to our studies are two LLs (M, 0) and (m, -1), between which the CNP of 4LG resides. At low field, the electron-like (m, -1) LL has higher energy than the hole-like (M, 0) LL. Taking Zeeman splitting into account, the $\nu=0$ state is an insulator with spin and valley polarization at $D=0$ (Fig. 4b); upon the application of $D$, this state crosses over smoothly into a layer-polarized insulator. This insulating phase at lower field ($B_\perp<22$ T) and over the entire range of accessible $D$ corresponds to phase I.

The insulating phase I, however, can transition into a metallic phase, if the hole-like (M, 0) LL surpasses the electron-like (m, -1) level, thus leading to counter-propagating edge states. Such a "band inversion" can be achieved by tuning either $B$ or $D$. First, at $D=0$, the different dispersions of these two LLs in $B_\perp$ lead to their crossing at sufficiently high magnetic fields. Fig. 4b plots the spin-split LL spectra $E(B_\perp)$ at $D=0$ near the crossing points, assuming a g-factor of 2. At $B_\perp=30$ T, the (m, -1, ↑) and (M, 0, ↓) LLs cross; the two valley-degenerate hole-like and two electron-like edge states disperse in opposite directions at the sample edge, giving rise to magnetic field-induced helical edge states that counterpropagate with opposite spin polarizations. This metallic phase corresponds to phase II with $\sim 4e^2/h$ conductance.

To examine the effect of $D$, we calculate the LL spectra at constant $B_\perp$ while varying the interlayer potential $\Delta$ (Fig. 4c-d). Here $\Delta$ is the actual potential difference between the top and bottom layers of 4LG, which, because of screening, is typically reduced by a factor of 5-7 from $Dd$, the experimentally imposed potential difference[27] ($d \sim 1$ nm is the thickness of 4LG). $\Delta$ breaks the inversion symmetry and lifts the valley degeneracy, and the different dispersion of valley-split LLs in $D$ gives rise to new crossing points.

Two representative $E(\Delta)$ spectra below and above $B_{\perp cp}=30$ T are shown in Fig. 4c-d, where the K and K' LLs are represented by solid and dashed lines, respectively. For instance, at $B_\perp=25$ T (Fig. 4c), the hole-like (M, 0, K', ↓) level is elevated above the electron-like (m, -1, K, ↑) when $\Delta$ exceeds 13 mV, giving rise to the conductive phase III, where there is only one "inverted LL", hence its conductance is $\sim 2e^2/h$. Further increase of $D$ above $\sim 20$ mV causes these two LLs to cross again, leading to the re-entrance of phase I. These two crossing points are labeled by hollow and solid circles in Fig. 4c, respectively. For $B_\perp>30$ T, the high magnetic field alone is sufficient to induce the "band inversion" at $\Delta=0$, and raising $\Delta$ gives rise to two distinct LL crossings. The first crossing occurs between (M, 0, K, ↓) and (m, -1, K', ↑), labeled by solid triangle in Fig. 4d, yielding a transition from phase II to III, where the number of hole-like LLs above the Fermi level is reduced from 2 to 1. The $\nu=0$ state reverts to a layer-polarized insulator when sufficient $\Delta$ is applied to fully valley-polarize the charge carriers, incurring (m, -1, K, ↑) crosses back with (M, 0, K', ↓), as indicated by the solid circle.

We reproduce the phase diagram in Fig. 2a by calculating $E(\Delta)$ at different $B_\perp$, and plot the crossing points in the $B_\perp$-$\Delta$ space (Fig. 4e), using the same symbols as in Fig. 4b-c to denote different LL crossings. The resulting phase diagram captures prominent features of the experimental data, most notably the sharp phase boundaries separating phase II from I and III. However, the single-particle model cannot account for the re-emergence of the low resistance state

(phase II) at large $B_\perp$ and intermediate $D$, suggesting the enhanced effect of interactions, e.g. exchange terms which favor spin-polarization.

To account for interaction effects, we introduce a minimal set of one-body and interaction terms, and perform Hartree-Fock (HF) calculations (see [28] for details). A typical phase diagram analysis is depicted in Fig. 4f. Comparing to the non-interacting result, it exhibits a richer structure and accounts for additional experimental features, including the re-entrance of high conduction phase for intermediate $D$, and smearing of the multi-critical point at $B_\perp \sim 30$ T and $D=0$ due to interaction-induced degeneracy lifting of the single particle levels. Two particularly robust phases emerge from the calculations: (a) a low-$B$ spin-singlet state ($S_z = 0$); it is largely determined by the one-body part of our model, and dominated by the competition between the one-body term ($\gamma_2$) that favors layer-coherence and $D$ that favors layer-polarization. Here 8 states of the form $|+; N\, s\rangle_a = \cos\frac{\theta_a}{2}|a; N\, s\rangle + \sin\frac{\theta_a}{2}|a+2; N\, s\rangle$ are occupied, where $a$=1, 2, $s$=↑, ↓ and $N$=0, 1 are layer, spin and orbital LL indices. The layer-polarization angles $\theta_a$ vary with $D$, from $\theta_a = \pi/2$ for $D \to 0$ to $\theta_a = 0$ in the high $D$ limit. Throughout this phase, both bulk and edge charge excitations are gapped and the system is an insulator, corresponding to phase I. (b). the second phase is a high-$B$ partially spin-polarized phase with $S_z = 2$ that is stabilized by interactions, where each orbital hosts four occupied single-electron states of the form $|a = 1, 3; N\, \uparrow\rangle$, $|+; N\, \downarrow\rangle_1$ and $|+; N\, \downarrow\rangle_2$. This ground-state corresponds to phase II, and has a bulk gap, but supports gapless helical edge modes protected by $S_z$ conservation. The resulting edge-dominated conductance $G \sim 2S_z e^2/h = 4e^2/h$ is compatible with the blue regions of Fig. 2a.

Interestingly, at large $B$ and moderately low $D$, a phase with $S_z=1$ appears, followed by the re-emergence of phase II with $S_z=2$ (Fig. 4f). These phases have 1 and 2 pairs of helical edge states, respectively. The precise boundaries of the less conducting $S_z=1$ state in the phase diagram (which appears compatible with the green regions in Fig. 2a) depend sensitively on the parameters of our model.

Finally, in narrower regions of the phase diagram, the HF analysis yields more complex states formed by a coherent superposition of various states $|a; N\, s\rangle$. These coherent states are zero temperature insulators; their continuous nature of the transitions into them suggest relatively small stiffnesses associated with the broken U(1) symmetries they host[38], hence small edge and bulk gaps and associated enhanced transport at non-vanishing $T$. The emergence of these broken symmetries can account, for example, for the enhanced transport observed at large $B$ and intermediate $D$ in the Corbino geometry. Further investigation of 4LG's phase diagram with refined parameters and measurements are warranted to fully understand the competing symmetries at the CNP.

**Acknowledgement** The experiments are supported by DOE BES Division under grant no. DE-SC0020187. Device fabrication is partially supported by the Center for Emergent Materials: an NSF MRSEC under award number DMR-1420451. Theoretical work was supported by the US-Israel Binational Science Foundation (Grant No. 2016130: GM, HAF, ES; Grant No.2018726: HAF, ES), by the NSF (Grant Nos. DMR-1506263, DMR-1914451, and ECCS-193640), and by the Israel Science Foundation (ISF) Grants No. 231/14 and 993/19 (ES). HAF acknowledges the support of the Research Corporation for Science Advancement through a Cottrell SEED Award. The authors acknowledge the hospitality and support of the Aspen Center for Physics (Grant No. PHY-1607611), where part of this work was done. GM is grateful to the Gordon and Betty Moore Foundation for sabbatical support at MIT, and the Lady Davis Foundation for sabbatical support at the Technion. Growth of hBN crystals was supported by the Elemental Strategy Initiative conducted by the MEXT, Japan and a Grant-in-Aid for Scientific Research on Innovative Areas "Science of Atomic Layers" from JSPS. We thank the groups of J. Hone and C. Dean for experimental advice on device fabrication.

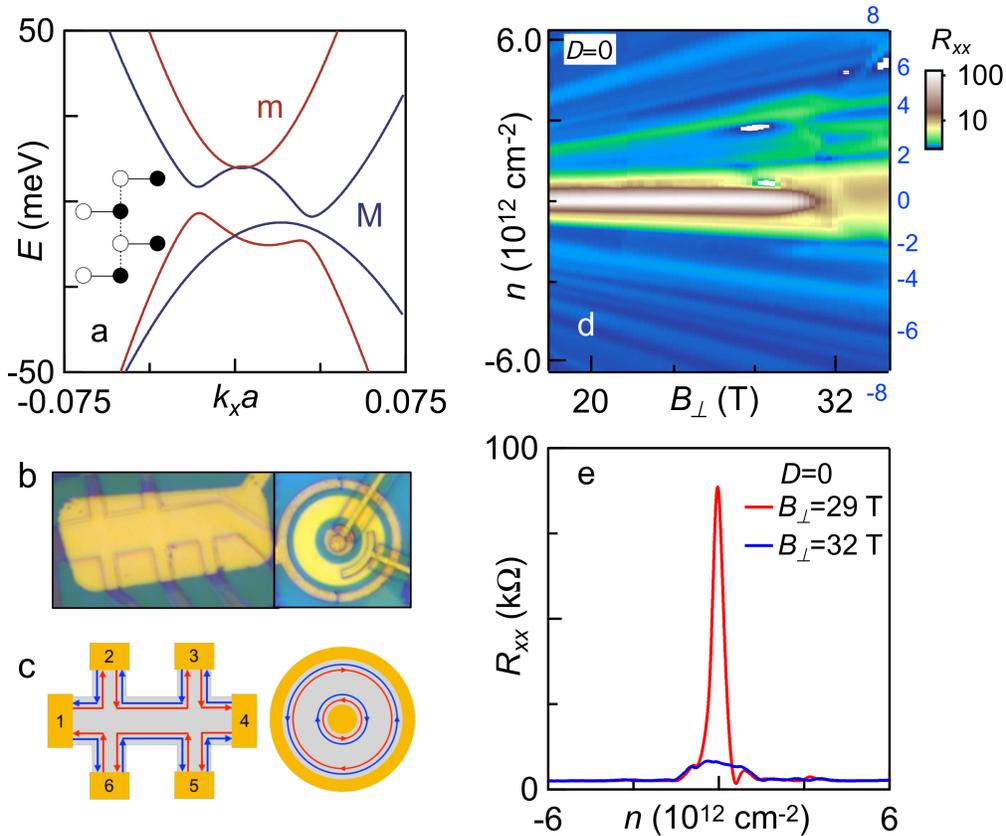

Fig. 1. (a) Low energy band structure $E(k_x)$ of 4LG in the absence of external fields. Red and blue lines denote BLG-like bands with light (m) and heavy (M) effective masses, respectively. (b-c) Schematics of Hall bar and Corbino devices, respectively, and propagation of 1 pair of helical edges therein. (d) Landau fan $R_{xx}(n, B_\perp)$ for a Hall bar device (H1) at $D=0$. The unit is k$\Omega$, in logarithmic scale. Blue numbers indicate filling factors. (e) $R_{xx}(n)$ at $D=0$ and $B_\perp=29$ and 32 T, respectively.

Fig. 2. (a) Electronic phase diagram $R_{xx}(D, B_\perp)$ at $n=D=0$, different phases are labeled I, II and III. The unit is kΩ. Dotted box indicates the region shown in Fig. 3a-b. (b) $R_{xx}(B_\perp)$ at $n=D=0$ at selected temperatures. (c-d) Temperature dependence of $R_{xx}(D)$ of $\nu=0$ state, at $B_\perp=25$ T and 34 T, respectively. Inset in (d): Zoom-in plot of $R_{xx}(D)$ at $T=0.3$ K, showing near quantization of phase I and II to $4e^2/h$ and $2e^2/h$, respectively.

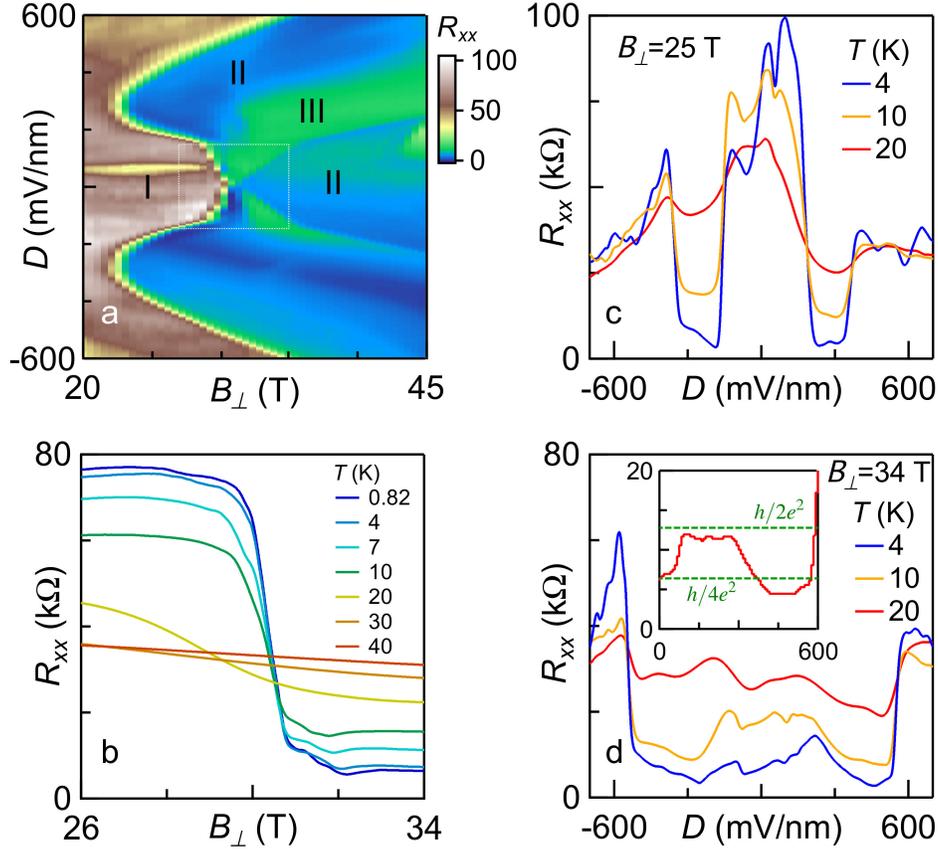

Fig. 3. (a) Conductance in unit of $e^2/h$ for a Corbino device (C1) as a function of $D$ and $B_\perp$, respectively. (b) Conductance contributed by edge states. (c) Phase diagram $R_{xx}(D, B_\perp)$ measured in tilt magnetic field at an angle $\theta=49°$, in unit of k$\Omega$. The red dotted curves outline the phase boundaries in Fig. 2a at $\theta=0$ and $B_{||}=0$. (d) $R_{xx}(D, B_\perp)$ at $n=D=0$ measured at different tilt angles from device H2. Inset: location of the quantum critical point in the plane of perpendicular ($B_\perp$) and total ($B_t$) magnetic field from 4 different devices. Solid line is a linear fit to the data.

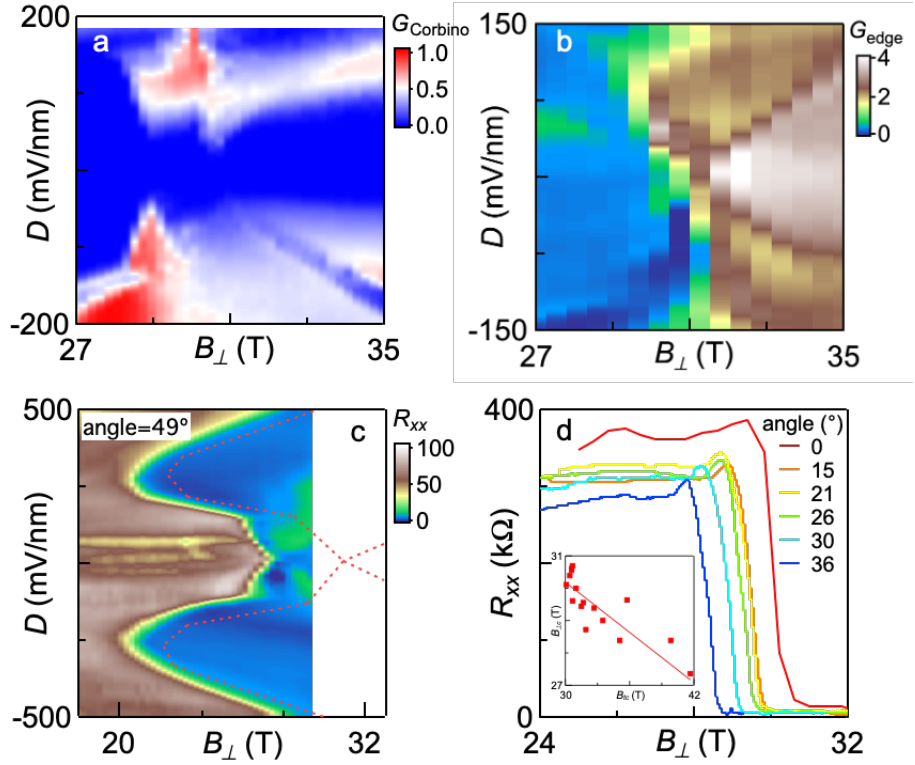

Fig. 4. (a) Spin- and valley-degenerate LL spectrum of 4LG at $D=0$. Red and blue lines denote electron and hole-like LLs, respectively. Numbers between the LLs indicate filling factors. LLs are labeled by heavy (M) or light (m) mass BLG-like band, orbital index and spin polarization. (b) Similar to (a), except for $25<B_\perp<45$T and each LL is spin split by a $g$-factor of 2. (c-d) LL energy vs. interlayer potential $\Delta$ at $B_\perp=25$ and 35 T, respectively. Circles and triangles indicate crossing points of electron and hole-like LLs that form the $\nu=0$ state. (e) Single particle phase diagram in $\Delta$-$B_\perp$ plane, using the same symbols for LL crossing points in (c-d). (f) Theoretical phase diagram by taking electronic interactions into account. PP1 and PP2 refer to two distinct ground states with partial spin polarization that support no conducting edges, with spin canting next-layer coherence and nearest-neighbor interlayer coherence, respectively[28].

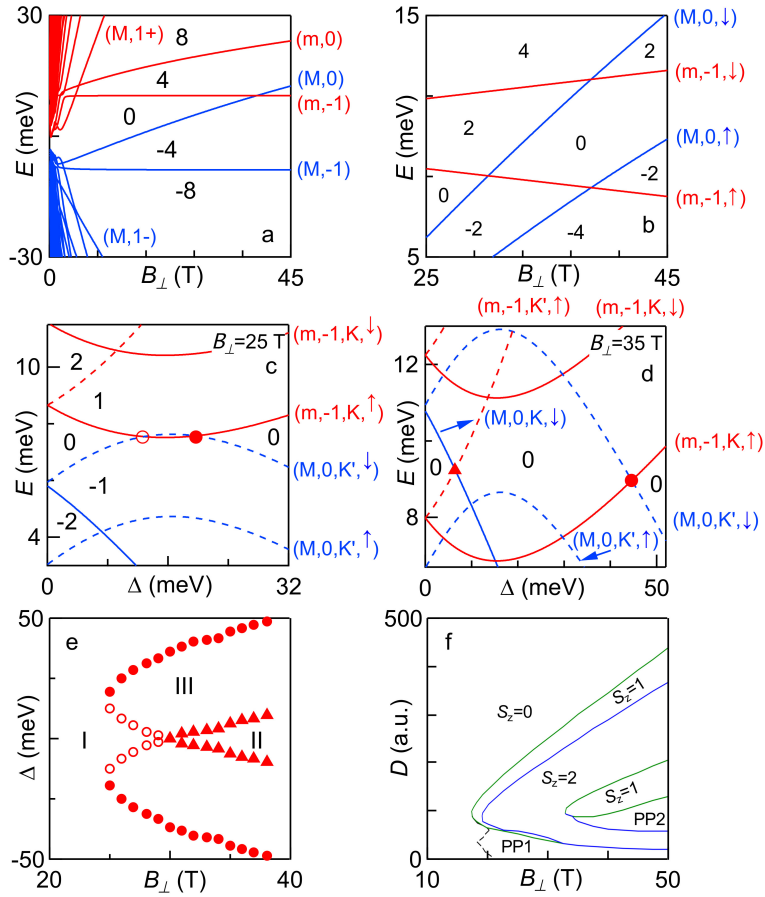